\documentclass{article}
\usepackage{graphicx} 

\usepackage{fancyhdr}
\usepackage{isomath}
\usepackage{mathtools} 
\usepackage{amsbsy}
\usepackage{amssymb}
\usepackage{amscd,amsfonts}
\usepackage{bigints}
\usepackage{graphicx}
\usepackage{verbatim}
\usepackage{euscript}
\usepackage{alltt}
\usepackage{stmaryrd}
\usepackage{relsize}
\usepackage{enumerate}
\usepackage{url}
\usepackage[stable]{footmisc}
\usepackage{breakurl}
\usepackage{hyperref}
\usepackage{comment}
\usepackage[numbers]{natbib}
\usepackage[font=small,labelfont=bf]{caption}
\usepackage[font=small,labelfont=bf]{subcaption}
\usepackage{float}
\usepackage{mwe}
\usepackage{algorithm}
\usepackage{algpseudocode}
\usepackage{xcolor}
\usepackage[super]{nth}
\usepackage{soul}

\DeclareGraphicsExtensions{.eps,.pdf}
\usepackage{amsmath}
\usepackage{amsbsy}
\usepackage{amssymb}
\usepackage{amscd}
\usepackage{amsfonts}

\newcommand{\R}{\mathbb R}

\newcommand{\K}{\mathbb{K}}

\newcommand{\beq}{\begin{equation}}
\newcommand{\eeq}{\end{equation}}
\newcommand{\beqs}{\begin{eqnarray}}
\newcommand{\eeqs}{\end{eqnarray}}
\newcommand{\beql}{\begin{equation} \label}
\newcommand{\half}{\frac{1}{2}}

\newcommand{\calA}{{\cal A}}

\newcommand{\calR}{{\cal R}}

\newcommand{\p}{\partial}

\usepackage[margin=1in]{geometry}

\newcommand{\dee}{\mathcal{D}}
\newcommand{\scl}{\mathcal{L}}

\date{}
\begin{document}
\title{Variational principle for a damped, quadratically interacting particle chain with nonconservative forcing}

\author{Amit Acharya\thanks{Department of Civil \& Environmental Engineering, and Center for Nonlinear Analysis, Carnegie Mellon University, Pittsburgh, PA 15213, email: acharyaamit@cmu.edu.} $\qquad$ Ambar N. Sengupta\thanks{Department of Mathematics, University of Connecticut, Storrs, CT 06269, email: ambarnsg@gmail.com.}}

\maketitle
\begin{abstract}
\noindent A method for designing variational principles for the dynamics of a possibly dissipative and non-conservatively forced chain of particles is demonstrated. Some qualitative features of the formulation are discussed.

\end{abstract}

\section{Introduction}
Starting from the work of Fermi-Pasta-Ulam-Tsingou \cite{fermi1955studies}, much has been understood about non-generic, non-thermalized behavior in the mechanics of discrete particle chains, e.g.~\cite{zabusky1965interaction, lax1968integrals, toda1970waves, moser1976three, gallavotti2007fermi}-\cite[and related Parts II, III, IV]{friesecke1999solitary}-\cite{smets1997solitary}-\cite[and Part II]{balk2001dynamics}, \cite{vainchtein2022solitary, truskinovsky2014solitary, schwetlick2007solitary, herrmann2010unimodal, pankov2019traveling, ingimarson2023long}. Such mechanical systems are considered to be idealized problems for understanding nonlinear dynamics of real solids. A variational description of such chains can help in the identification of interesting solutions like periodic orbits and solitary waves, and their stability, e.g. \cite{smets1997solitary, schwetlick2007solitary, herrmann2010unimodal,  pankov2019traveling}. Our recent work \cite{acharya2023action} outlines a general setup for constructing variational principles for general anholonomically constrained, possibly dissipative Newtonian particle systems. Here, we apply those ideas to the problem of a special class of particle chains to enable future work on the subject from this point of view. We have chosen the case of at most quadratic force interactions in order to explicitly write out the action functional involved.

We note here that the existence of variational action principles for systems with dissipation and non-conservative forces is a surprising and novel result; it has generally been believed that dissipative systems do not admit variational action principles. Moreover, Hamilton's principle does not yield an initial value problem as its Euler Lagrange equations and side conditions but a boundary value problem whose final-time boundary condition is unknown for solving initial value problems - see \cite{galley2013classical} for a discussion. In this connection, we also note that constructing an action principle whose first variation gives a certain set of dissipative equations is not the same as incorporating a Rayleigh dissipation function to the Lagrangian of an associated conservative system and obtaining the Euler-Lagrange equations of the resulting functional; a typical dissipative force in the Rayleigh formalism with Rayleigh function $\calR(\dot{q})$ is given by $ - \frac{\p \calR}{\p \dot{q}}$ which is not the same as the variational derivative $\frac{\delta \int_0^T \calR(\dot{q}(t)) \, dt}{\delta q}$.

\section{Dual variational principles for ODE systems: the general formalism}
We briefly review our strategy for developing action principles for particle systems \cite{acharya2023action}. Consider a (primal) system of ODE for a vector of degrees of freedom $t \mapsto U(t) \in \R^s$ for some $s \geq 1$, and $F$ is a given function of its arguments with $t$ being time:
\begin{equation}\label{eq:abs_ODE}
    \calA \, \dot{U} - F(U,t) = 0; \qquad \qquad U(0) = U^{(0)},
\end{equation}
and $\calA$ is a constant matrix. Treating the above system as constraints for extremizing an arbitrarily chosen objective functional $\int_0^T H(U,t) \, dt$ and introducing \emph{dual} Lagrange multiplier functions $t \mapsto D(t) \in \R^s$, a pre-dual functional is defined as
\[
\widehat{S}_H[U,D] = \int^T_0 \scl_H(U,\dee,t) \, dt - \calA \, U^{(0)} \cdot D(0); \qquad  \dee: = (D, \dot{D}); \quad  \scl_H(U,\dee, t) := - \calA \, U \cdot \dot{D} - D \cdot F(U,t) + H(U,t),
\]
where we include only those `boundary' terms that allow inclusion of information from the primal problem. We now require the function $H$ to be such that it allows for solving the algebraic equation $\frac{\p \scl_H}{\p U}(U,\dee,t) = 0$ for $U$ in terms of $(\dee,t)$, i.e. it allows the definition of a function, the \textit{dual-to-primal} (DtP) mapping, $U = U^{(H)}(\dee,t)$ such that
\begin{equation}\label{eq:dldu}
    \frac{\p \scl_H}{\p U} \left( U^{(H)}(\dee,t), \dee, t \right) = 0
\end{equation}
is satisfied in some neighborhood of the space of $(\dee, t)$.

The desired action functional, \textit{for any $H$ with the above property}, is then defined by substituting the function $U^{(H)}$ in the pre-dual functional $\widehat{S}$ for $U$:
\[
S_H[D] = \int_0^T \scl_H\left( U^{(H)}(\dee(t),t), \dee(t), t \right) \, dt - \calA \, U^{(0)} \cdot D(0), \qquad D \ \mbox{arbitrarily specified at} \ t = T,
\]
and it can be checked that its Euler-Lagrange (E-L) equations and implied b.c. are simply the primal system \eqref{eq:abs_ODE} with $U$ replaced by $U^{(H)}(\dee,t)$.

\section{A dual action functional for a forced particle chain with damping and quadratic interaction forces}
We apply this machinery to a 1-d chain of $N$ particles that have the same mass $m > 0$, $d \geq 0$ is the damping coefficient, $x \mapsto K(x) \in \R^N $ is a  quadratic function (allowing for long-range interactions) characterized by the given constants $C \in \R^N$, $A \in \R^{N \times N}$, and $B \in \R^{N \times N \times N}$, $B$ symmetric in the last two indices, and $t \mapsto f(t) \in \R^N$ is a prescribed forcing function of time (not necessarily the gradient of a scalar potential in $x$):
\begin{equation}\label{eq:chain_ode}
    \begin{aligned}
       &  m \dot{v}_i + d v_i + K_i(x) - f_i(t) = 0 \\
       &  \dot{x}_i - v_i  = 0\\
        & \dot{x}_i(0) = x_i^{(0)};  \qquad \dot{v}_i(0) = v_i^{(0)}\\
        & K_j(x) :=K_j(\bar{x}) + A_{jr} (x_r - \bar{x}_r) + \frac{1}{2} B_{jrs} (x_r - \bar{x}_r) (x_s - \bar{x}_s),
    \end{aligned}
\end{equation}
where the interaction force $K(x)$ is at most a quadratic polynomial in the particle positions, expanded about an arbitrarily chosen `base' configuration $\bar{x}$ to be discussed further below.

Now consider a pre-dual functional with a shifted quadratic form for the auxiliary potential $H$:
\begin{equation}\label{eq:pre_dual}
    \begin{aligned}
        \widehat{S}[x,v, \gamma, \lambda] & = \int^T_0 - v_i m \dot{\lambda}_i + \lambda_i d v_i  + \lambda_i K_i(x) - \lambda_i f_i - x_i \dot{\gamma}_i - \gamma_i v_i + \frac{1}{2} c_x |x - \bar{x}|^2 + \frac{1}{2} c_v |v - \bar{v}|^2\, dt \\
        & \qquad \qquad - \lambda_i(0) m v_i^{(0)} - \gamma_i(0) x_i^{(0)},
    \end{aligned}
\end{equation}
where the shift is by an arbitrarily chosen `base state' $(t \mapsto (\bar{x}(t), \bar{v}(t))$.
Define
\begin{equation*}
    U = (x,v); \qquad \qquad \dee = (\gamma, \lambda, \dot{\gamma}, \dot{\lambda}); \qquad \qquad \scl(U, \dee; \bar{U}) = \mbox{integrand of \eqref{eq:pre_dual}}
\end{equation*}
and solve for $U$ in terms of $(\dee, \bar{U})$ from the equations
\[
\frac{\p \scl}{\p U}(U, \dee; \bar{U}) = 0
\]
to obtain the system
\begin{subequations}\label{eq:mapping}
    \begin{align}
        \frac{\p \scl}{\p x_i} &: \qquad \lambda_j( A_{ji} + B_{jri} (x_r - \bar{x}_r)) - \dot{\gamma}_i + c_x \delta_{ir} (x_r - \bar{x}_r) = 0 \nonumber\\
        \Longrightarrow & \ \qquad c_x  \K_{ir}\big|_\lambda (x_r - \bar{x}_r) = \dot{\gamma}_i - \lambda_j A_{ji};  \qquad \mbox{where} \quad \K_{ir}\big|_\lambda := \delta_{ir} + \frac{1}{c_x} \lambda_j B_{jir}\\
        \frac{\p \scl}{\p v_i} &: \qquad - m \dot{\lambda}_i + d \lambda_i - \gamma_i + c_v (v_i - \bar{v}_i) = 0 \nonumber\\
        \Longrightarrow &  \ \qquad c_v (v_i - \bar{v}_i) = \gamma_i + m \dot{\lambda}_i - d \lambda_i
    \end{align}
\end{subequations}
Substituting \eqref{eq:mapping} in \eqref{eq:pre_dual}, we obtain
\begin{equation}\label{eq:pre_dual_2}
    \begin{aligned}
        \widehat{S}[x,v, \gamma, \lambda] & = \int_0^T (v_i - \bar{v}_i) (-m \dot{\lambda}_i + d \lambda_i - \gamma_i) + \frac{1}{2} c_v (v_i - \bar{v}_i) (v_i - \bar{v}_i) \, dt \\
        & \quad + \int_0^T \bar{v}_i (- m \dot{\lambda}_i + d \lambda_i - \gamma_i) - \bar{x}_r \dot{\gamma}_r + \lambda_i K_i\big|_{\bar{x}} - \lambda_i f_i \, dt\\
        & \quad + \int_0^T \half \lambda_j B_{jri} (x_r - \bar{x}_r) (x_i - \bar{x}_i) + \half c_x \delta_{ri} (x_r - \bar{x}_r) (x_i - \bar{x}_i) - (x_i - \bar{x}_i) \left( \dot{\gamma}_i - \lambda_j A_{ji} \right)\, dt\\
        & \quad - \lambda_i(0) m v_i^{(0)} - \gamma_i(0) x_i^{(0)}\\
        & = \int_0^T - \half c_v (v_i - \bar{v}_i) (v_i - \bar{v}_i) - \half c_x \K_{ir} (x_r - \bar{x}_r)(x_i - \bar{x}_i)\, dt \\
        & \quad + \int_0^T \bar{v}_i (- m \dot{\lambda}_i + d \lambda_i - \gamma_i) - \bar{x}_r \dot{\gamma}_r + \lambda_i K_i\big|_{\bar{x}} - \lambda_i f_i \, dt\\
        & \quad - \lambda_i(0) m v_i^{(0)} - \gamma_i(0) x_i^{(0)}.
    \end{aligned}
\end{equation}
Finally, using \eqref{eq:mapping} in \eqref{eq:pre_dual_2} we obtain a \emph{dual} functional
\begin{equation}\label{E:Sdual}
    \begin{aligned}
        S[\gamma, \lambda] & = - \half \int_0^T \frac{1}{c_v} \big|\gamma + m \dot{\lambda} - d \lambda \big|^2 + \frac{1}{c_x} ( \dot{\gamma} - A^T \lambda) \cdot \K \big|_\lambda ^{-1} \left( \dot{\gamma} - A^T \lambda \right) \, dt\\
        & \quad + \int_0^T - \bar{v} \cdot \left(\gamma + m \dot{\lambda} - d \lambda \right) - \bar{x} \cdot \dot{\gamma} + \lambda \cdot K\big|_{\bar{x}} - \lambda \cdot f \, dt\\
        & \quad - \lambda(0) \cdot m v^{(0)} - \gamma(0) \cdot x^{(0)}\\
        & \quad \mbox{with} \ \lambda(T) = 0, \quad \gamma(T) = 0 \ \mbox{without loss of generality}.
    \end{aligned}
\end{equation}

We make the following observations:
\begin{itemize}
    \item Since the E-L equations of $S$ are guaranteed to be the system in \eqref{eq:chain_ode} (with the DtP mapping in place), if the functions $(\bar{x}, \bar{v})$ are actual solutions to \eqref{eq:chain_ode}, then the DtP mapping \eqref{eq:mapping} suggests that $t \mapsto (\lambda(t), \gamma(t)) = (0,0)$ is an extremal of $S$. Thus, at least for dual functionals designed with base states that are `close to' actual (un)stable solutions of the primal problem, it is reasonable to expect solutions (extremals) to exist for the dual functional.
    \item Why should it be possible to solve an initial-value-problem by prescribing final-time boundary conditions as well in time, especially when the primal problem has a unique solution for prescribed initial conditions? This is because the DtP mapping equations \eqref{eq:mapping} show that the mapped primal functions at the final time, $(x(T), v(T))$, depend on the rates $(\dot{\lambda}(T), \dot{\gamma}(T))$, and specifying $(\lambda(T), \gamma(T))$ leaves enough freedom in the dual problem to adjust the rates to satisfy the demands of recovering the correct `nearby' primal solution - this has been discussed and demonstrated in \cite[Sec.~7]{action_2}-\cite{KA1}.
    \item The dual E-L equations is a coupled system of 2nd-order, 2-point boundary value problems in the dual variables $(\lambda, \gamma)$ whose ellipticity is governed by  the matrix
    \begin{equation*}
        \begin{bmatrix}
            \frac{m^2}{c_v} \mathbb{I} & 0\\
            0 & \frac{1}{c_x} \K\big|_\lambda^{-1}
        \end{bmatrix}
    \end{equation*}
    which is positive definite as long as $\K$ remains positive definite. Thus, in a neighborhood of $t \mapsto (\lambda(t), \gamma(t)) = (0,0)$ corresponding to the primal trajectories $(\bar{x}, \bar{v})$, the dual problem is elliptic and therefore, if $(\bar{x}, \bar{v})$ is a primal solution, then the corresponding dual solution $t \mapsto (\lambda(t), \gamma(t)) = (0,0)$ is neutrally stable - this can also be seen by calculating the second variation of the dual functional $- S$ about the orbit $t \mapsto (\lambda(t), \gamma(t)) = (0,0)$ and noticing that it is non-negative (the Lagrangian of the dual action $S$ is concave in $\dee$ and hence it is the (local) maximization of $S$ that is relevant for defining primal solutions through the DtP mapping).
    \item The dual functional results in second order equations for the dual variables. Seeking dual extremals in the class of periodic functions in time - with the period as an additional variable that does not evolve (see \cite[Sec.~3, point 5.]{acharya2023action}) provides a seamless way of seeking periodic solutions in time (e.g.~`breathers') for a nonlinear initial value problem, which is otherwise a non-trivial question even at the formal procedural level and computationally, even more so.
    \item As is well-known, the existence of constants of motion for nonlinear chains plays a significant role in their analysis and understanding. If the Lagrangian of the dual functional is strictly monotone in $\dot{D}$, then an equivalent Hamiltonian description of the dual problem can be written down \emph{for each chosen $H$}, thus resulting in the possibility of defining several constants of motion for the dual problem. Since each solution of the dual problem is associated with a solution to the primal problem, this seems to suggest, at least superficially, that such constants of motion can be associated with primal orbits as well (cf.~\cite{acharya2023action}). This feature of our formulation is interesting, and awaits further study.
\end{itemize}

\section{Conclusion}

In this paper we have described a novel method, developed more extensively in our recent work \cite{acharya2023action}, to construct a variational principle from a given dynamical equation that could involve nonholonomic constraints and dissipative forces. The variational principle is framed in terms of certain dual variables that are related to the given primal variables and an appropriate new objective function $H$. We have applied our method to a system of $N$ particles experiencing an applied force that is quadratic in the position coordinates, an additional time-dependent force, and a dissipative force, linear in the velocity, as well. We  obtain in (\ref{E:Sdual}) an exact form of the variational action functional for this system in terms of the dual variables. We made some observations concerning the dual  action functional and possible implications for the existence of constants of motion.

\section*{Acknowledgments}
This work was supported by the Simons Pivot Fellowship grant \# 983171 and NSF OIA-DMR grant \# 2021019.

\bibliographystyle{alphaurl}\bibliography{quad_particle}
\end{document}